\documentclass[preprint,onecolumn,aps]{revtex4-1}

\usepackage{graphicx}
\usepackage{multirow}
\usepackage{float}
\usepackage{bm}
\usepackage{subfig} 
\usepackage{caption}
\usepackage{amsmath}
\usepackage{amssymb}

\begin{document}


\title{On the bending of rectangular atomic monolayers along different directions: an ab initio study}

\author{Shashikant Kumar}
\affiliation{College of Engineering, Georgia Institute of Technology, Atlanta, GA 30332, USA}

\author{Phanish Suryanarayana}
\email{phanish.suryanarayana@ce.gatech.edu}
\affiliation{College of Engineering, Georgia Institute of Technology, Atlanta, GA 30332, USA}


\begin{abstract}
We study the bending of rectangular atomic monolayers along different directions from first principles. Specifically, choosing the phosphorene, GeS, TiS$_3$, and As$_2$S$_3$ monolayers as representative examples, we perform Kohn-Sham density functional theory calculations to determine the variation in transverse flexoelectric coefficient and bending modulus with the direction of bending. We find that while the flexoelectric coefficient is nearly isotropic, there is significant and complex anisotropy in bending modulus that also differs between the  monolayers, with extremal values not necessarily occurring along the principal directions. In particular, the commonly adopted orthotropic continuum plate model with uniform thickness fails to describe the observed variations in bending modulus for GeS, TiS$_3$, and As$_2$S$_3$. We determine the direction-dependent effective thickness for use in such continuum models. We also show that the anisotropy in bending modulus is not associated with the rehybridization of atomic orbitals. 
\end{abstract}


\maketitle

\section{Introduction}

Over the past two decades, crystalline atomic monolayers --- dozens have now been synthesized \cite{vaughn2010single, singh2022synthesis, akhtar2017recent, shi2015recent, lv2015transition} and thousands  have  been predicted to be stable from first principles Kohn-Sham density functional theory (DFT) calculations \cite{haastrup2018computational, zhou20192dmatpedia} --- have been the subject of intense research \cite{mas20112d, butler2013progress, geng2018recent}. This is due to their interesting and exotic mechanical \cite{balendhran2015elemental, zhang2018recent}, electronic \cite{zhou2018monolayer, fei2015giant, hsu2017topological, dai2016group}, and optical properties \cite{marjaoui2022strain, bouziani2022two, zhou2022giant}, which are typically muted or non-existent in their bulk counterparts. The most common lattice structures among these 2D materials are honeycomb and rectangular. While the properties of honeycomb monolayers are generally found to be  isotropic \cite{lee2009elastic, sun2021first, falin2017mechanical, lee2008measurement}, i.e., independent of in-plane direction,  significant anisotropy is common in rectangular monolayers \cite{vannucci2020anisotropic, Gao2022anisotropy, liu2021highly, wang2015design}.

The responses/properties of atomic monolayers under mechanical deformations are important in several technological applications, including flexible electronics \cite{yoon2013highly, salvatore2013fabrication}, nanoelectromechanical devices \cite{tan2020out, sazonova2004tunable}, nanocomposites \cite{novoselov2012two, qin2015lightweight}, wearable mechanical sensors \cite{yang2015tactile, zhang2019skin, cheng2015stretchable}, and single-photon emitters \cite{rosenberger2019quantum, peng2020creation}. This has motivated a number of studies on the mechanical properties of atomic monolayers, both experimental \cite{han2020ultrasoft, lee2008measurement, cooper2013nonlinear, bertolazzi2011stretching, lindahl2012determination, ruiz2011softened} and theoretical/DFT \cite{kudin2001c, liu2007ab, wei2009nonlinear, lu2009nonlinear, Shenderova2002Carbon, huang2006thickness}; as well as on their response to mechanical deformations, both experimental \cite{guinea2010energy, dai2019strain, meng2013hierarchy, ong2012engineered, peng2020strain, zhu2013strain} and theoretical/DFT \cite{han2020effect, taheri2018effects, zhu2015thermal, deng2018strain, liu2017strain, ghorbani2013strain, kripalani2018strain, gui2008band}. These efforts have generally focused on tensile deformations, since bending requires sophisticated  experiments with high accuracy in measurements \cite{akinwande2017review}; and ab initio DFT simulations are computationally intensive,  scaling cubically with system size, which makes them nonviable at practically relevant bending curvatures \cite{kumar2020bending}. Indeed, such studies can be performed using computationally cheaper alternatives such as tight binding \cite{verma2016directional, zhang2011bending, tapaszto2012breakdown} and classical force fields \cite{arroyo2004finite, koskinen2010approximate, cranford2009meso, cranford2011twisted, roman2014mechanical, liu2011interlayer, xu2010geometry, sajadi2018size, qian2020multilayer}. However, these  methods typically lack the resolution required to study nanoscale systems such as monolayers, as is evident by the significant scatter in the reported  bending moduli values for even elemental monolayers, e.g., 0.8 to 2.7 eV for graphene \cite{arroyo2004finite, sajadi2018size}, and 0.4 to 38 eV for silicene \cite{roman2014mechanical, qian2020multilayer}.

In recent work, cyclic+helical symmetry-adapted DFT calculations \cite{sharma2021real, ghosh2019symmetry} have been used  to  compute the bending moduli for forty-four atomic monolayers \cite{kumar2020bending} as well as the transversal flexoelectric coefficient --- measures the rate of change of the out-of-plane dipole moment with curvature, which arises due to the bending-induced strain gradient across the thickness \cite{codony2021transversal} --- for fifty-four atomic monolayers \cite{kumar2021flexoelectricity}, along their principal directions.  It has been found that atomic monolayers with honeycomb lattice, i.e., group IV monolayers, transition metal dichalcogenides (TMDs), group III monochalcogenides, and group IV dichalcogenides, have bending modulus and flexoelectric coefficient values that do not vary between the principal directions, whereas those with rectangular lattice, i.e., group V monolayers, group IV monochalcogenides, transition metal trichalcogenides (TMTs), and group V chalcogenides, have significantly different bending modulus but nearly same flexoelectric coefficient values along the principal directions. These and previous ab initio studies have however not considered  the bending of  monolayers along directions that are different from the two principal directions, which provides the motivation for the current work. 

In this work, we study the bending of rectangular atomic monolayers along different directions using Kohn-Sham DFT. Specifically, choosing the phosphorene, GeS, TiS$_3$, and As$_2$S$_3$ monolayers as representative examples --- each has been synthesized, and belongs to a notable  monolayer group that has a rectangular lattice structure --- we investigate the  variation in transverse flexoelectric coefficient and bending modulus with the bending direction. We find that while the flexoelectric coefficient is nearly isotropic, there is significant and complex anisotropy in the bending modulus that also differs between the monolayers. For each of the monolayers, we determine the direction-dependent effective thickness to be used in orthotropic continuum plate models. We also study the correlation between the underlying electronic structure and the direction-dependent bending modulus. 

\section{Methods}

\begin{figure}[h!]
	\centering
		\includegraphics[width=0.8\textwidth]{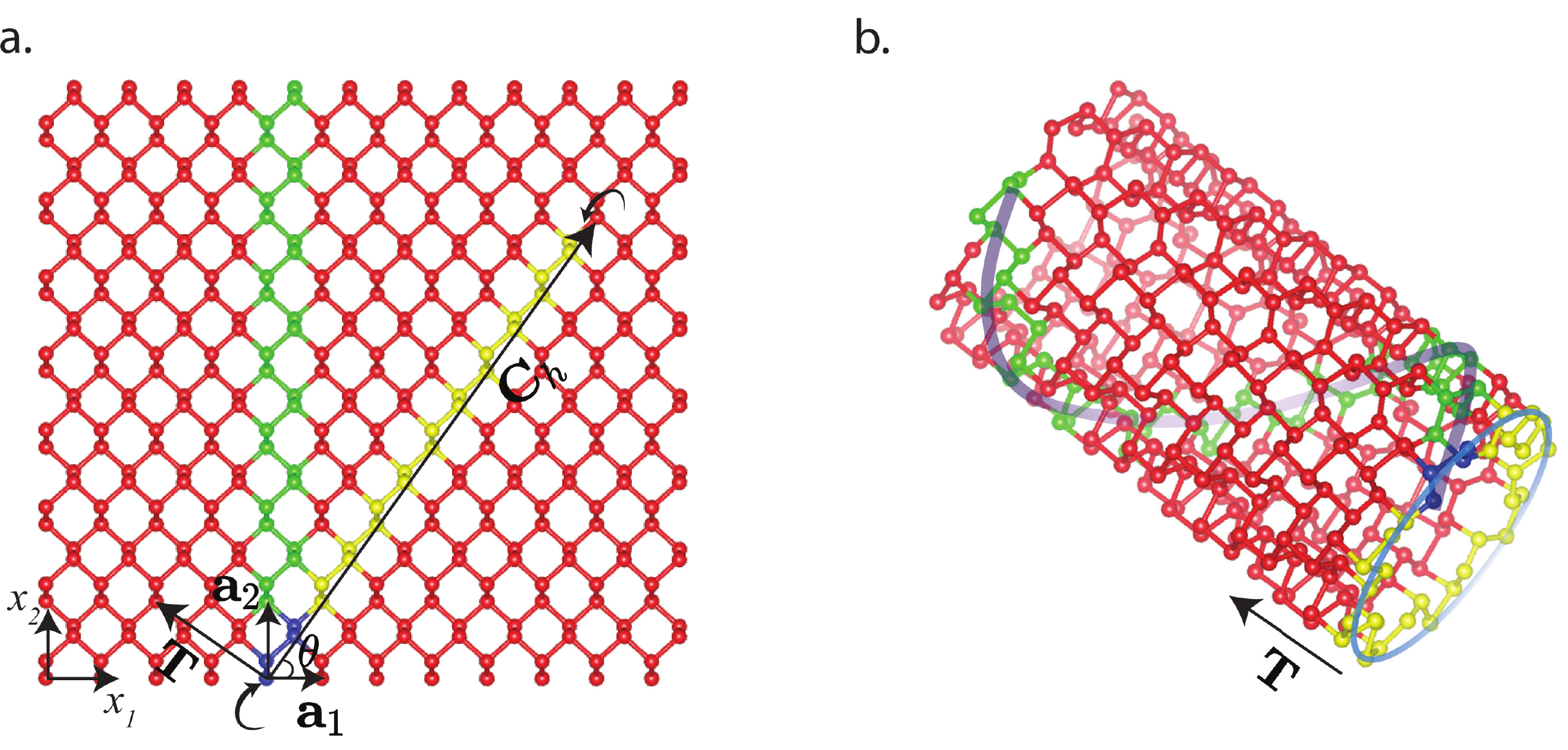}	
		\caption{Illustration showing the nanotube generated by bending of a phosphorene monolayer along an arbitrary direction. $\mathbf{a}_1$ and $\mathbf{a}_2$ are the lattice vectors for the monolayer, $\theta$ represents the direction of bending, $\mathbf{C}_h$ is the  chiral vector, and $\mathbf{T}$ is the translation vector. The atoms in the fundamental domain/unit cell are shown in blue color, with  cyclic and helical symmetry-related images shown in yellow and green colors, respectively. The structural model has been generated using VESTA \cite{momma2011vesta}. }
		 \label{Fig:illlustration}
\end{figure}

We perform Kohn-Sham DFT calculations using the real-space electronic structure code SPARC \cite{xu2021sparc, ghosh2017sparc2, ghosh2017sparc1}. Specifically, we simulate the bending of the selected atomic monolayers using the Cyclix-DFT feature \cite{sharma2021real}, which has been well tested in a number of physical applications \cite{bhardwaj2022strain, bhardwaj2022elastic, bhardwaj2021torsional, bhardwaj2021torsionalmoduli, codony2021transversal, kumar2021flexoelectricity, kumar2020bending, sharma2021real}. In particular, edge-related effects are removed by considering the nanotube obtained by rolling the monolayer along a certain direction, with the nanotube's radius chosen to be equal to the desired bending radius of curvature  \cite{banerjee2016cyclic, ghosh2019symmetry, sharma2021real}.   The cyclic and helical symmetry of the resulting nanotube is then exploited to reduce the computations to the fundamental domain ---  possesses same number of atoms as the monolayer's periodic unit cell  --- thereby significantly accelerating the calculations \cite{sharma2021real}, enabling efficient simulations in the practically relevant low-curvature limit.  See Fig.~\ref{Fig:illlustration} for an illustration of the bending of the phosphorene monolayer along an arbitrary direction, with  the resulting chiral nanotube having only four atoms in its fundamental domain. Note that due to geometrical constraints,  in order for it to be possible to roll a rectangular monolayer into a nanotube with arbitrary chiral index,  the ratio of squares of the lattice constants needs to be  an integer \cite{dovesi2017crystal17}.  We ensure this in the current study by imparting a small axial strain along one of the principal directions. Indeed, since the flexoelectric coefficient and bending modulus involves taking  differences in quantities between varying bending curvatures,  significant error cancellation is to be expected, whereby the applied strain is not expected to noticeably impact the results.

For a given bending direction, we calculate the transversal flexoelectric coefficient $\mu_{\rm T}$ using the relation \cite{codony2021transversal, kumar2021flexoelectricity, codony2022comment}:
\begin{equation} \label{Eq:FlexoCoefficient}
\mu_{\rm T} = \frac{\partial }{\partial \kappa} \left(  \frac{1}{A} \int_{\Omega} (r-R_{\rm eff}) \rho(\mathbf{x}) \, \mathrm{d \mathbf{x}}  \right)\,,
\end{equation}
where $1/\kappa$ is the radius of the nanotube, $A$ is the nanotube's cross-sectional area within the fundamental domain $\Omega$, $r$ is the radial coordinate at  $\mathbf{x}$, $R_{\rm eff}$ is the radial ionic centroid, and $\rho(\mathbf{x})$ is the ground state electron density. In particular, the flexoelectric coefficient is computed by employing a numerical approximation to the derivative in Eq.~\ref{Eq:FlexoCoefficient}, i.e., the quantity in brackets --- referred to as the radial polarization --- is evaluated at multiple curvatures near the curvature at which the flexoelectric coefficient needs to be computed, and the corresponding curve-fit is used to approximate the derivative.

For a given bending direction, we calculate the bending modulus $D$ by fitting data to the expression:
\begin{align}
    \label{Eq:BM}
        \mathcal{E} (\kappa) &= \mathcal{E}_0 + \frac{1}{2} D A \kappa^2 \,,
\end{align}
where $\mathcal{E}(\kappa)$ is the ground state energy of the nanotube with radius $1/\kappa$.

In all simulations, we employ the Perdew-Burke-Ernzerhof (PBE) \cite{perdew1986accurate}  exchange-correlation functional, and ONCV \cite{hamann2013optimized} pseudopotentials from the SPMS \cite{spms} collection. The computed lattice constants for the monolayers (Supplementary Material) are in good agreement with both experimental \cite{woomer2015phosphorene, li2012role,island2015tis3, siskins2019highly} and theoretical studies \cite{zhou20192dmatpedia, haastrup2018computational, balendhran2015elemental, kang2015mechanical, gomes2015phosphorene, singh2014computational}, verifying the accuracy of the chosen exchange-correlation functional and pseudopotentials.  We consider ten bending directions for each of the  selected atomic monolayers: phosphorene, GeS, TiS$_3$, and As$_2$S$_3$. In order to simulate mildly bent sheets, i.e., calculate quantities corresponding to the low curvature limit, we choose bending curvatures: $0.15 \leq \kappa \leq 0.25$ nm$^{-1}$, commensurate with experimental studies for bending  \cite{lindahl2012determination}. All numerical parameters in Cyclix-DFT, including the real-space and Brillouin zone grid spacings, vacuum in the radial direction, and relaxation tolerances for cell/atom are chosen such that the computed flexoelectric coefficient and bending modulus values are accurate to within $0.01e$ and 1\%, respectively. In practice, this requires the computed ground state energy to be converged to within $10^{-5}$ Ha/atom, which is necessary to capture the extremely small differences, particularly those arising during the computation of the bending modulus. 

We note that without the use of the  cyclic+helical symmetry-adapted framework \cite{sharma2021real}, many of the simulations needed here would have been tremendously expensive, if not impossible, e.g., the As$_2$S$_3$ system with $\kappa=0.15$ nm$^{-1}$  and bending along $\theta = 61.2$ degrees has a total of $583,520$ electrons in the unit cell for periodic boundary conditions, which is well beyond the reach of traditional DFT formulations/implementations. This reduces to only 56 electrons in the symmetry-adapted framework, which is identical to the number in standard periodic unit cell calculations for the monolayer. 
 
\section{Results and Discussion}
We now present results of the aforedescribed Kohn-Sham DFT bending simulations for the phosphorene, GeS, TiS$_3$, and As$_2$S$_3$ atomic monolayers. Additional details regarding the simulation data and results presented/discussed here can be found in the Supplementary Material.

\begin{figure}[h!]
	\centering
		\includegraphics[width=0.95\textwidth]{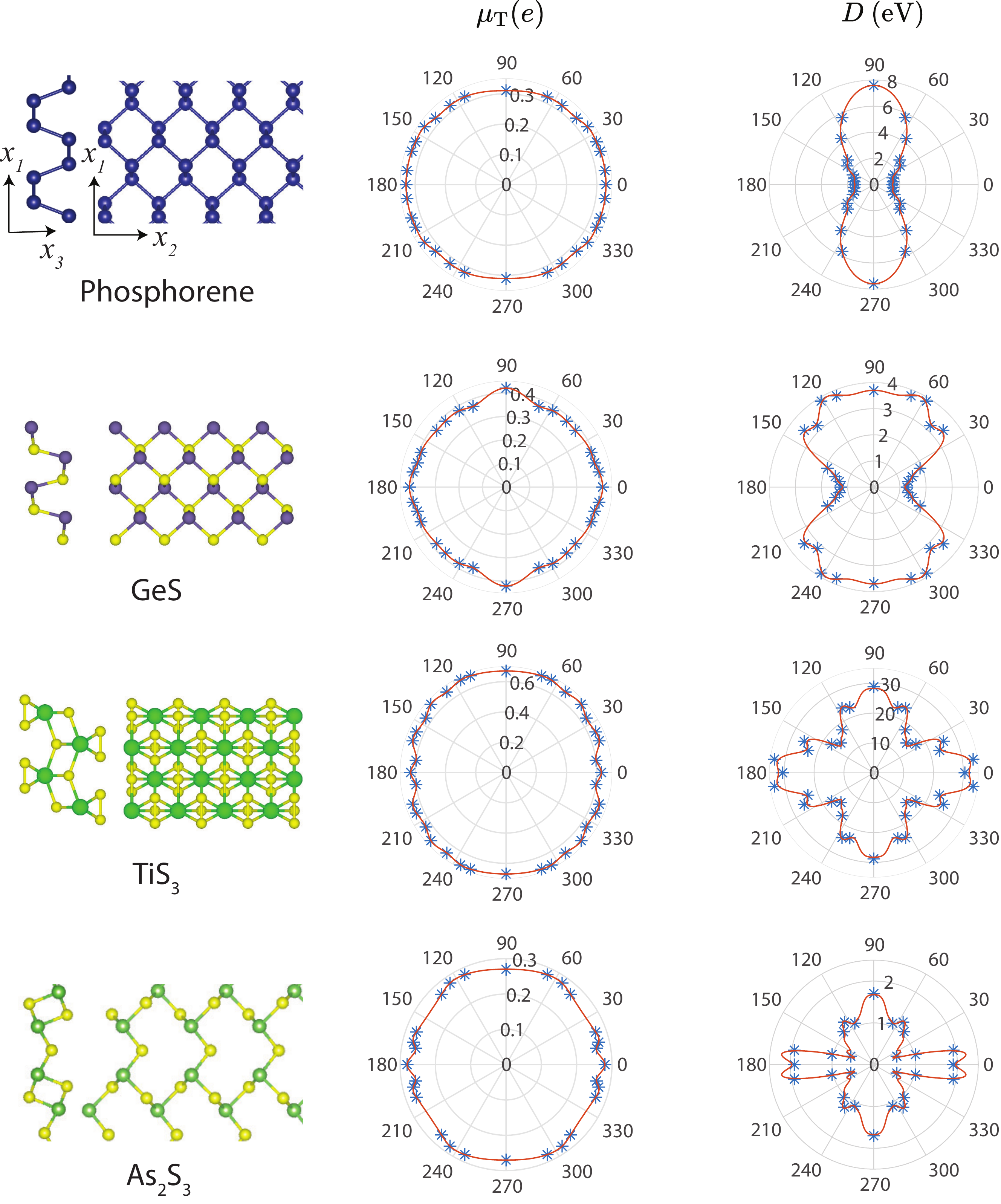}	\caption{Transversal flexoelectric coefficient ($\mu_{\rm T}$) and bending modulus ($D$) along different directions for the phosphorene, GeS, TiS$_3$, and As$_2$S$_3$ monolayers,  as computed from Kohn-Sham DFT simulations. The markers represent the data points and the solid line represents a curve fit of the form: $f(\theta) = c_0 + \sum_{n=1}^{7} [c_n \cos(2n\theta) + s_n \sin(2n\theta)]$. }
		 \label{Fig:results}
\end{figure}

In Fig.~\ref{Fig:results}, we present the variation in the values of the transversal flexoelectric coefficient $\mu_{\rm T}$ and bending modulus $D$ with the direction of bending. On the one hand, we observe that the flexoelectric coefficient is nearly independent of direction, indicating that it is isotropic for the chosen monolayers.  On the other hand, the bending modulus is  noticeably affected by the bending direction, resulting in significant and complex variation that also differs between the different monolayers (likely due to the different underlying lattice and electronic structures), indicating that it is highly anisotropic. In particular, the ratio of maximum to minimum bending modulus  for the phosphorene, GeS, TiS$_3$, and As$_2$S$_3$ monolayers is 5.3, 3.1, 2.3, and 3.3, respectively, with the maximum and minimum values not occurring  along the principle directions for the GeS, TiS$_3$, and As$_2$S$_3$ monolayers, i.e., all monolayers but phosphorene.

The values of the flexoelectric coefficient and bending modulus  along the principle directions are in excellent agreement with those reported in Refs.~\cite{kumar2021flexoelectricity, kumar2020bending}, which also employ DFT calculations with the same exchange-correlation functional, i.e., PBE. Comparisons for other directions cannot be made due to the unavailability of experimental/DFT studies in literature, as also noted in the introduction. In the case of phosphorene, there has been a recent study  on its direction-dependent bending modulus using the tight binding approximation \cite{verma2016directional}. While there is reasonable agreement in the qualitative features between   Ref.~\cite{verma2016directional} and the current work,  there are significant quantitative differences, with values deviating by as much  as 2 eV. These differences can be attributed to the approximate nature of tight binding methods, particularly when compared to Kohn-Sham DFT, further highlighting the need for ab initio calculations in such studies. 

\begin{figure}[h!]
	\centering
		\includegraphics[width=1.0\textwidth]{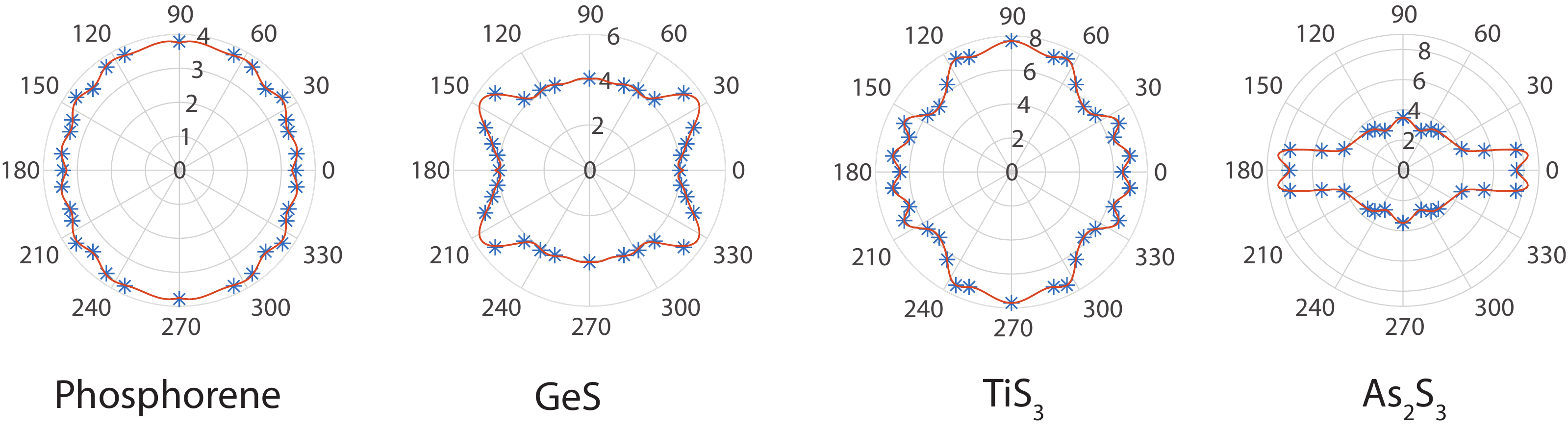}	\caption{Direction-dependent effective thickness (\AA) for the phosphorene, GeS, TiS$_3$, and As$_2$S$_3$ monolayers, for use in the orthotropic continuum plate model. The markers represent the data points and the solid line represents a curve fit of the form: $f(\theta) = c_0 + \sum_{n=1}^{7} [c_n \cos(2n\theta) + s_n \sin(2n\theta)]$.}
		 \label{Fig:results_thickness}
\end{figure}

It is common to employ an orthotropic continuum plate model  for the higher-scale analyses of atomic monolayers, e.g., vibrations and instabilities \cite{liu2012stochastically, adali2012variational, murmu2009buckling, liu2021highly, wei2014superior, ren2022prediction, zhang2011bending, garcia2008imaging}.  In this model, the out-of-plane bending modulus is related to the in-plane Young's modulus $Y$ and Poisson's ratio $\nu$ through the relation \cite{reddy1999theory}: 
\begin{align}
    \label{Eq:D_Y_continuum}
        D(\theta) = \frac{Y(\theta)t^3}{12\left(1 - \nu(\theta)\nu(\theta+\pi/2)\right)} \,,
\end{align}
where $t$ is the thickness of the plate. To check the validity of this model in the current context, given that the thickness of monolayers is not well-established \cite{huang2006thickness, shearer2016accurate, Rickhaus2020thickness}, we first calculate their direction-dependent Young's modulus and Poisson's ratio in the flat configuration, again using the SPARC electronic structure code \cite{xu2021sparc, ghosh2017sparc2, ghosh2017sparc1}. Next, we substitute the computed bending modulus, Young's modulus, and Poisson's ratio into Eq.~\ref{Eq:D_Y_continuum} to determine the effective thickness as a function of direction, i.e., $t(\theta)$,  the results for which are presented in Fig.~\ref{Fig:results_thickness}. We observe that while the effective thickness is nearly independent of direction for phosphorene, there is significant anisotropy for the other monolayers, suggesting the failure of the continuum model when using a constant thickness for these systems. In particular, the direction-dependent effective thickness determined here can be used in such continuum models for higher-scale analyses. Note that the effective thickness for phosphorene reported by Ref.~\cite{verma2016directional}  is significantly more anisoptropic than the one here, again a likely consequence  of Ref.~\cite{verma2016directional}  using the more approximate tight binding methods. 

\begin{figure}[h!]
	\centering
		\includegraphics[width=\textwidth]{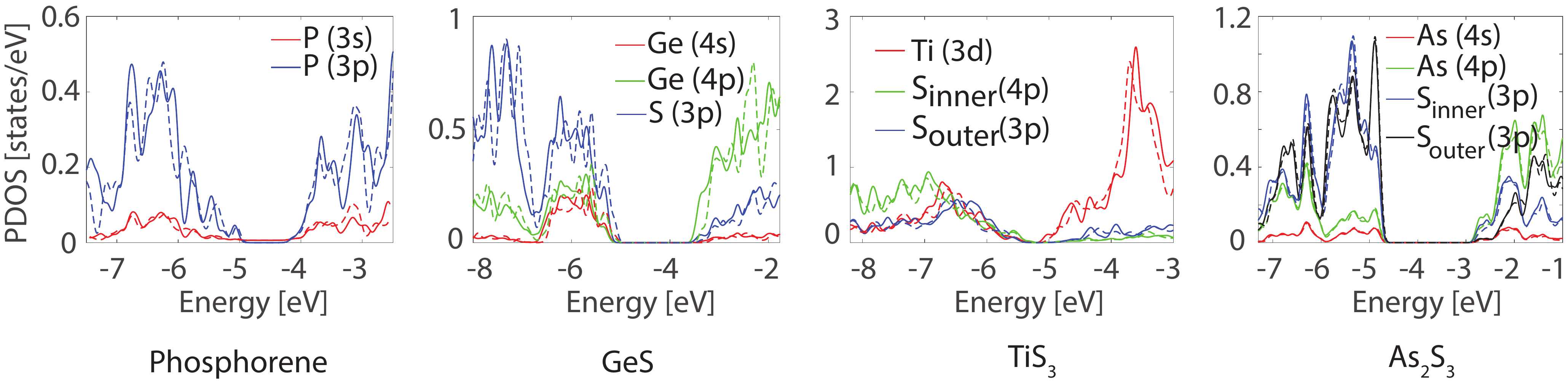}	\caption{Atomic orbital projected density of states (PDOS) for bent phoshorene, GeS, TiS$_3$, and As$_2$S$_3$ monolayers ($\kappa$ = 0.25 nm$^{-1}$), with directions of bending corresponding to those along which the bending modulus is maximum (solid line) and minimum (dashed line).}
		 \label{Fig:PDOSplots}
\end{figure}

To get further insight into the observed   anisotropy in the bending modulus, we calculate the atomic orbital projected density of states (PDOS) for the bent monolayers, i.e., nanotubes. In Fig.~\ref{Fig:PDOSplots}, we plot the PDOS so obtained for the directions along which the monolayer  has the largest and smallest bending modulus. We observe that while there are significant differences at the level of the magnetic quantum number resolved PDOS (Supplementary Material), the curves at the level of the angular quantum number (i.e., summed over the  magnetic quantum number, for each principal and angular quantum number) are similar for both bending directions. Given that we are considering small bending curvatures that do not significantly  change the relative orientation of the atoms from the flat sheet configuration, the PDOS results suggest that the anisotropy in the bending modulus is not a consequence of the rehybridization of orbitals, but rather due to the the structure-dependent directional weakening of bonds.

\section{Concluding remarks}
In this work, we have studied the bending of four rectangular atomic monolayers: phosphorene, GeS, TiS$_3$, and As$_2$S$_3$, along different directions, from first principles. In particular, we have performed Kohn-Sham  DFT calculations to determine the variation in transverse flexoelectric coefficient and bending modulus with bending direction. We have found that the flexoelectric coefficient is nearly isotropic, whereas the bending modulus has significant and complex anisotropy that also differs between the monolayers,  with the maximum and minimum values not necessarily occurring along the  principal directions. In particular, the orthotropic continuum plate model --- commonly  employed in literature for atomic monolayers --- fails to describe the observed variations in bending modulus for GeS, TiS$_3$, and As$_2$S$_3$. We have determined the direction-dependent effective thickness that can be used in such continuum models, which has applications in higher-scale vibrational and instability analyses. We have also found that the anisotropy in bending modulus is not a consequence of the rehybridization of atomic orbitals, but rather due to the structure-dependent directional weakening of bonds. The bending of bilayers/multilayers  and heterostructures presents itself as a worthy subject of future research. 

\section*{Acknowledgements}

The authors gratefully acknowledge the support of the US National Science Foundation (CAREER-1553212).

\providecommand{\newblock}{}

\end{document}